\documentclass[12pt]{article}
\usepackage{latexsym,amsmath,amssymb,amsfonts,amsthm,amstext,amscd}
\usepackage{DraTex,AlDraTex}

\newcommand{\ie}{\textit{i.e.}$\,$}
\newcommand{\Real}{\mathbb{R}}

\newcommand{\cF}{\mathcal{F}}
\newcommand{\vel}{\mathbf{v}}\newcommand{\cel}{\mathbf{c}}
\newcommand{\xel}{\mathbf{x}}

\newcommand{\wel}{\mathbf{w}}

\newcommand{\B}{\mathbf{B}}

\newcommand{\id}{\operatorname{id}}

\newcommand{\der}{\operatorname{der}}

\newcommand{\bt}{\begin{tabular}{c}}
\newcommand{\et}{\end{tabular}}
\newcommand{\ba}{\begin{array}{c}}
\newcommand{\ea}{\end{array}}

\def\beginf{\begin{flushright}}
\def\ef{\end{flushright}}

\swapnumbers
       \theoremstyle{definition}
\newtheorem{odef}{Definition}[section]
\newtheorem{exe}[odef]{Exercise}

\newtheorem{oexm}[odef]{Example}
\newtheorem{onotation}[odef]{Notation}

      \theoremstyle{plain}
\newtheorem{thm}[odef]{Theorem}

\theoremstyle{remark}

\begin{document}\title{Concepts of relative velocity}
\author{Zbigniew Oziewicz
\\{\small Universidad Nacional Aut\'onoma de M\'exico}\\{\small Facultad de Estudios Superiores Cuautitl\'an}\\{\small C.P. 54714 Cuautitl\'an Izcalli, Apartado Postal \# 25, Estado de M\'exico}\\oziewicz@unam.mx\\and\\William S. Page\\{\small 1219 Thousand Islands Parkway}\\{\small Mallorytown, Ontario, Canada K0E 1R0}\\bill.page@newsynthesis.org}
\date{March 29, 2011}\maketitle

\hyphenation{Na-tu-ral ex-pe-ri-men-tal o-pe-ra-tor pa-ra-me-te-ri-zing}

\begin{abstract} The central concept of the theory of relativity is the \textbf{relativity} of velocity. The velocity of a material body is not an intrinsic property of the body; it depends on a free choice of reference system. Relative velocity is thus reference-dependent, it is not an absolute concept. We stress that even zero-velocity must be relative. Every reference system possesses its own zero-velocity relative only to that particular reference system. Does the theory of relativity formulated in terms of relative velocities, with many zero-velocities, imply the Lorentz isometry group? We discuss the many relative spaces of Galileo and Poincar\'e, as quotient spaces.\end{abstract}

\medskip

\noindent\textbf{Physics and Astronomy Classification Scheme (PACS) 1999/2000.} 03. special relativity \medskip

\tableofcontents

\pagestyle{myheadings}\markboth{\quad\hfill
Oziewicz and Page: Concepts of relative velocity\hfill }{\emph{Oziewicz and Page: Concepts of relative velocity}\hfill\quad}

\section{Einstein's relative velocity is ternary}\hspace{5mm} In the present paper relativity means the historical term `special relativity', where we drop `special' because the theory of relativity, in our understanding, is \textit{coordinate-free}.

The Lorentz isometry is frequently presented as a transformation of coordinates. However the concept of an isometry does not exist without first specifying the metric tensor in a coordinate free manner, \ie an isometry does not exist without a scalar product.

Starting from the metric tensor Fock derived the following particular Lorentz-boost transformation [Fock 1955, 1959, 1961, 1964 \S 10 and \S 16; Jackson 1962, 1975 \S 11.3]. Our question is: Of what exactly is this a transformation?
\begin{gather}\gamma\equiv\frac{1}{\sqrt{1-\frac{\vel^2}{c^2}}},\qquad
\xel'=\xel+\frac{\gamma^2}{\gamma+1}\frac{(\vel\cdot\xel)}
{c^2}\vel-\gamma\vel t,\label{F}\\\
t'=\gamma\left(t-\frac{\vel\cdot\xel}{c^2}\right)\qquad\Longleftrightarrow\qquad
\frac{\vel\cdot\xel}{c^2}=t-\frac{t'}{\gamma}\label{F2}\end{gather}

Fock's expression needs the scalar product $\vel\cdot\xel.$ One can ask where is this scalar product? In the spacetime? or in a three-dimensional space?

For two-dimensional spacetime \eqref{F}-\eqref{F2} collapses to Einstein's expression below \eqref{E} [Einstein 1905], however we must stress that the naive generalization of Einstein's coordinate transformation to more dimensions is \textit{not} an isometry,
\begin{gather}\ba \text{isometry}\\x'=\gamma\left(x-vt\right)\ea\qquad\nRightarrow\qquad\ba\text{not isometry}\\\xel'=\gamma\left(\xel-\vel t\right)\ea\label{E}\end{gather}

We are interested in the precise definition and interpretation of all symbols in \eqref{F}-\eqref{F2}.
What it is the meaning of the symbol $\vel,$ generating transformation of what? It is the relative velocity of what body relative to what reference? If $\vel$ is a vector, on spacetime or on some space, then \eqref{F} implies that also $\xel$ must be a vector, and not just a set of coordinates. If the symbol $\xel$ denotes a vector, then \eqref{F} implies the vanishing of the Grassmann bivector
\begin{gather}\eqref{F}\quad\Longrightarrow\quad(\xel'-\xel)\wedge\vel=0.\label{biv}\end{gather}
Where it is the above bivector \eqref{biv}, in four-dimensional spacetime? or in some three-dimensional space?

Inserting \eqref{F2} into \eqref{F} allows us to express the velocity $\vel$ in terms of a vector $\xel-\xel',$ this solves \eqref{biv} explicitly. Still there is only an implicit $(\xel-\xel')$-dependence because of the Lorentz factor $\gamma=\gamma(\vel)$ is $\vel$-dependent,
\begin{gather}\vel=\left(1+\frac{1}{\gamma(\vel)}\right)\left(\frac{\xel-\xel'}{t+t'}\right).\label{v}\end{gather} In the Galilean limit, $c\longrightarrow\infty,$ and $t'=t,$ the above expression of relative velocity collapses to the widely accepted expression. Now we can insert \eqref{v} into the Lorentz factor $\gamma$ \eqref{F}, and this allows us to express $\gamma$ in terms of the scalar product $(\xel-\xel')^2,$
\begin{gather}\frac{\vel^2}{c^2}=\left(1+\frac{1}{\gamma}\right)\left(1-\frac{1}{\gamma}\right)\qquad\Longrightarrow\qquad
\gamma=\frac{1+\frac{1}{c^2}\left(\frac{\xel-\xel'}{t+t'}\right)^2}
{1-\frac{1}{c^2}\left(\frac{\xel-\xel'}{t+t'}\right)^2}\label{gamma}\end{gather}
Finally inserting $\gamma=\gamma((\xel-\xel')^2,\ldots)$ \eqref{gamma} into \eqref{v} gives the desired operational expression of the velocity in terms of the vector $\xel-\xel',$ \ie the expression ready for experimental measurement of relative velocity within the Lorentz group-relativity,
\begin{gather}\vel=2\left\{1+\frac{1}{c^2}
\left(\frac{\xel-\xel'}{t+t'}\right)^2\right\}^{-1}\left(\frac{\xel-\xel'}{t+t'}\right).\label{Urbantke}\end{gather}

The above explicit expression \eqref{Urbantke} for the relative velocity parameterizing the Lorentz transformation \eqref{F}-\eqref{F2}, was derived by Urbantke in another way, using reflections, \ie involutory isometries,  [Urbantke 2003, p. 115, formula (7)]. Previously Ungar derived the same expression using gyration [Ungar 2001, p. 348, Theorem 11.16],
\begin{gather}\vel=\frac{\xel'}{t'}\circleddash\text{gyr}\left[\frac{\xel'}{t'},\frac{\xel}{t}\right]\frac{\xel}{t}.\label{U}\end{gather}

The above expression, Ungar  \eqref{U}, and Urbantke \eqref{Urbantke}, can be adopted as the \textit{definition} of relative velocity in Einstein's special relativity in terms of directly measured quantities. However, isometry implies a more general definition. There is also the question of who is actually measuring the relative velocity according to the above formulae \eqref{Urbantke}?

The expression \eqref{Urbantke} is an easy consequence of the Lorentz isometry transformation \eqref{F}-\eqref{F2}, and at least it should always be presented jointly with the Lorentz transformations. Just to verify, insertion of \eqref{Urbantke} into \eqref{F}-\eqref{F2}, gives an identity.

\begin{exe} Using symmetry \eqref{M} below, one can derive other expressions for relative velocity that are equivalent to Ungar's and Urbantke's expression \eqref{Urbantke}.\end{exe}

Do we like this definition of relative velocity? The actual physical concept of relative velocity has not yet even been discussed.

The velocity of one material (or massive) body is always relative to another body, or, we could say that the velocity of a body is always relative to a free \textit{choice} of reference system. In fact relativity theory is a theory of massive reference systems.

Most definitions of velocity, including \eqref{Urbantke}-\eqref{U}, are \textbf{obscured} by imposing a coordinate system. These coordinate systems contain implicit, hidden or incomplete, obscure information about the material bodies involved.

If we do not define or make precise the meaning of material body, then the concept of velocity is meaningless. The exact concept of material body is crucial for understanding the concept of velocity. Georg Hegel (1770--1831) wrote: \textit{no motion without matter}.

Light is massless and therefore can not be considered to define a reference system. The same applies to cosmic background radiation. \textbf{The velocity of light must \textit{not} be considered a primary concept of the theory of relativity.} A similar opinion is shared by [Paiva and Ribeiro 2005] who claim that \textit{special relativity does not depend on electromagnetism}\footnote{Relativity need not postulate the velocity of light. However it does seem to need an invertible metric tensor on spacetime and this metric tensor is involved in Maxwell electromagnetism. In particular the metric determines the constitutive properties of `empty space', \ie $\varepsilon$ and $\mu$ which are related to the speed of light by Maxwell's theory.}.

\section{Lorentz transformation without metric? No.}\hspace{5mm} The Lorentz group is a symmetry group of the metric tensor on spacetime (this tensor is sometimes strangely called the `interval' in [Zakharov 2006]). What does this mean? We must stress \begin{center}\textbf{no metric} $\quad\Longrightarrow\quad$ \textbf{no Lorentz group of isometries.}\end{center}

Our questions about the transformation of coordinates \eqref{F}-\eqref{F2} are:
\begin{enumerate}
\item Where is the metric tensor? Where is the scalar product on spacetime?\item What is the interpretation of each symbol in \eqref{F}?\item What is the physical meaning of the transformation vector $\vel$ if we understand that isometries are generated by a bivector, not by a vector?\end{enumerate}

How could it be nowadays that so many textbooks\footnote{For example. In Wolfgang Rindler's \textit{Essential Relativity}, Springer 1969, 1977, the Lorentz transformation is derived on pages 32-33, but the concept of the metric is introduced on page 62.} of `special' relativity still present the Lorentz group in terms of coordinate transformations \eqref{F}-\eqref{F2}, without even mentioning the metric \textit{tensor} on spacetime? We think that the omission of the metric tensor is a crime, and presenting this tensor in the diagonal form, something that it is possible only in particular basis, while tensors are basis-free, has brought even more misunderstanding.

The one textbook interpretation of \eqref{F}-\eqref{F2} is that there are two reference systems and that both observe an event $e.$ However time can not be stopped. An event as a point on spacetime manifold is not observable, it does not exist in nature. What can be observed is the word line of an event, \ie `event' must be a life history of a material particle, or better a time-like vector field $E.$

Textbooks interpret $\{\xel,t\}$ and $\{\xel',t'\},$ as two different observers in two different reference systems.

In fact there are three actors: street s, bus b, and eagle e. What is the `position' vector $\xel$? Let's say that it is the `position' vector of the eagle as seen from the street reference system, $\xel(s,e).$ Analogously from the bus moving with respect to the street, $\xel'=\xel(b,e).$ The quantities in \eqref{F} depend on the motion of the eagle, and therefore the relative velocity of the bus relative to the street a priori should be a function of three variables, viz. $\vel(s,b,e).$ It is not obvious how this `relative' velocity can be independent of the eagle. In fact we should suppose that the eagle is a third reference system. Thus the relative velocity $\vel$ in \eqref{Urbantke} is ternary, not binary.

The Lorentz-boost isometric transformation deduced by Fock, \eqref{F}-\eqref{F2}, presupposes the following symmetry of the scalar product,
\begin{gather}-(ct)^2+\xel^2=-(ct')^2+\xel'^{\,2}=-1.\label{M}\end{gather}
Fock started his deduction of \eqref{F} by first exhibiting the metric tensor $g.$ This metric \textit{tensor} is still implicit in the scalar form \eqref{M}. We need to incorporate this metric tensor $g$ explicitly. We will then re-derive \eqref{F}-\eqref{F2} in a coordinate-free manner below.

A material reference system can be modeled in terms of a time-like vector field on space-time. This was proposed by Minkowski in 1908: a material reference system is a normalized time-like vector, a monad, and not some basis = tetrad. Within this philosophy the domain of the Lorentz isometry transformation \textbf{must} be the all vectors tangent to spacetime. Tensors and vectors (a vector is a tensor) are coordinate-free. A transformation of vectors induces the transformation of all tensors. Lorentz transformation operates on all vectors and tensors in a coordinate-free manner. All tensors are $GL$ and Lorentz-covariant, but one tensor, the metric tensor will remain Lorentz-invariant. The Lorentz transformation is an isometry, and it is coordinate-free, when acting on vectors.

Let $S$ be a time-like vector field, $S^2\equiv g(S\otimes S)=-1.$ The associated differential one-form `$-gS$' is said to be an $S$-proper-time form, and \begin{gather}(-gS)S=1.\end{gather} Therefore, $s\equiv S\otimes(-gS),$ is an idempotent, $s^2=s,$ $sS=S,$ and $(\id-s)$ is also idempotent.

For any vector field $E$ we have the following coordinate-free identity\begin{gather}E=sE+(\id-s)E.\end{gather} Here, $sE=-(S\cdotp E)S,$ is time-like, and it is orthogonal to the space-like $(\id-s)E,$
\begin{gather}S\cdotp(\id-s)E=S\cdotp\{E+(S\cdotp E)S\}=0,\quad
E^2=(sE)^2+((\id-s)E)^2.\end{gather}

Let a time-like vector field $E$ represent the eagle, $E^2=-1,$ with an associated idempotent, $e\equiv E\otimes(-gE),$ $e^2=e.$

In what follows the time-like vector $S$ represents the reference street, with an associated idempotent $s^2=s.$ Let moreover a time-like vector field $B,$ $B^2=-1,$ represent the bus with associated idempotent $b^2=b.$ The eagle seen from the street and from the bus is as follows
\begin{gather}E=sE+(\id-s)E=bE+(\id-b)E.\end{gather}

\begin{onotation}[Eagle observed from street and from bus] We introduce the following notation-conventions,
\begin{gather}sE=ctS\quad\text{\ie}\quad ct\equiv-S\cdotp E\equiv -g(S\otimes E),\\
bE=ct'B\quad\text{\ie}\quad ct'\equiv-B\cdotp E\equiv-g(B\otimes E),\\
\xel\equiv(\id-s)E,\qquad\xel'\equiv(\id-b)E,\\
(\xel'-\xel)\wedge S\wedge B\equiv 0,\quad S\cdotp\xel=0,\quad B\cdotp\xel'=0.\end{gather}\end{onotation}
The difference of the position vectors, $(\xel'-\xel)$ must be co-planar with a plane $S\wedge B.$
With above notation, the metric symmetry \eqref{M} implies that the eagle must be represented as a time-like vector field $E^2=-1.$

\begin{onotation}[Eagle observing street and bus]\label{Eobserving} Alternatively one can suppose that an eagle e is observing the motion of the bus relative to the street,
\begin{gather}\begin{aligned}S&=eS+(\id-e)S=ct\,E\,+\xel,\\B&=eB+(\id-e)B=ct'E+\xel'.\end{aligned}\label{observing}\\(\xel'-\xel)\wedge E\wedge S\wedge B=0,\quad E\cdotp\xel=0=E\cdotp\xel'.\label{colinear}\end{gather}\end{onotation}
Here we see that the difference of the position vectors, $(\xel'-\xel)$ must be within a three-dimensional volume $E\wedge S\wedge B.$

\section{Lorentz transformation without bivectors? No.}\hspace{5mm} We are interested in the concept of the velocity of the bus relative to the street, \ie the velocity as measured by the eagle. Why should the eagle be the involved in this relative velocity concept?

We need to define the Lorentz-boost isometry-transformation of coordinate-free vectors.

The Lie algebra of the Lie group of isometries $\text{Aut}(g)\simeq O(1,3),$ coincides with the vector space of Grassmann bi-vectors inside Clifford algebra. A Minkowski bivector $P\wedge Q$ generates an isometry
\begin{gather}P\wedge Q\quad\hookrightarrow\quad L_{P\wedge Q}\in O(1,3),\\
\text{street}\quad\xrightarrow{\quad\text{Lorentz-boost}\quad}\quad\text{bus},\\S\quad\xrightarrow{\quad L_{\text{bivector}}\quad}\quad B=L_{\text{bivector}}S.
\end{gather}

Here we arrive at what we consider the essence of special relativity theory, that isometries are generated by bivectors, not vectors.

\begin{thm}[Isometry-link problem (Oziewicz 2007)] Given a massive three-body system in terms of three time-like normalized vectors $\{E,S,B\}.$ Let a space-like Minkowski vector $\wel$ be observed by $E$ \ie $E\cdotp\wel=0.$ Then the Lorentz-boost-link equation for the unknown $\wel,$ $$L_{E\wedge\wel}\,S=B\quad\text{with}\quad E\cdotp\wel=0,$$ has a unique solution, $\gamma\frac{\vel}{c}\equiv\wel=\wel(E,S,B).$\end{thm}

\begin{onotation}[Unbounded speed] The vector $\wel\equiv\gamma\vel/c$ is said to be unbounded, whereas $\vel<\cel$ is bounded by the light speed. If $\vel$ approaches the light speed, then $\gamma\rightarrow\infty,$ and $|\wel|\rightarrow\infty.$\end{onotation}

We say that the unbounded velocity $\wel$ of the bus $B$ relative to the street $S$ is observed by the eagle $E.$ All isometric relative velocities are ternary [Oziewicz 2007, 2009; Celakoska 2008; Celakoska and Chakmakov 2010].

Elsewhere we derived the following general expression for the isometry generated by bivector $E\wedge\wel$ with $E\cdotp\wel=0$ [Oziewicz 2006-2009],
\begin{gather}L_{E\wedge\wel}S=S-\{(\gamma-1)E\cdotp S-\wel\cdotp S\}E-\left(E\cdotp S-\frac{\wel\cdotp S}{\gamma+1}\right)\wel\end{gather}

Let $\wel$ be unbound velocity of the bus $B$ relative to street $S$ as measured by eagle $E,$  $B=L_{E\wedge\wel}S.$ Using Notation \ref{Eobserving} we have
\begin{gather}eL_{E\wedge\wel}S=(\wel\cdotp S-\gamma E\cdotp S)E\quad\Longleftrightarrow\quad t'=\gamma\left(t+\frac{\vel\cdot\xel}{c^2}\right).\end{gather}

This proves the Fock transformation \eqref{F2}, and clarifies that \begin{itemize}
\item The scalar product $\vel\cdotp\xel$ is a scalar product of vectors in spacetime - not space!
\item The actual relative velocity $\vel$ is eagle $E$-dependent [Oziewicz 2007].
\end{itemize}

\begin{proof} Now we will prove the Fock expression \eqref{F2}. Using Notation \ref{Eobserving} we have
\begin{gather}\xel'\equiv(\id-e)L_{E\wedge\wel}S=S+(E\cdotp S)E-\left(E\cdotp S-\frac{\wel\cdotp S}{\gamma+1}\right)\wel,\\\xel\equiv(\id-e)S=S+(E\cdot S)E,\\\xel'-\xel=\frac{\gamma^2}{\gamma+1}\frac{\vel\cdotp\xel}{c^2}\vel-\gamma\vel t.\end{gather}\end{proof}

\section{Space is not physical reality}\hspace{5mm} All considerations above take place in spacetime. In spacetime there is a unique zero velocity. In order to introduce many zero relative velocities, each zero for each reference system, we must consider the mathematical conventions of Galileo and Poincar\'e concerning many relative spaces as quotient spaces.

Galileo stressed in 1632 that all velocities are relative and our everyday experience tells us the same thing. One can not discover one's own motion without looking outside for another reference system. The theory of relativity  by Galileo 1632, and by Poincar\'e 1902, is all about the concept of relative velocity.

\begin{quotation}\ldots treatises on mechanics do not clearly distinguish between what is experiment, what is mathematical reasoning, what is convention, and what is hypothesis.

There is no absolute space, and we only conceive of relative motion; and yet in most cases mechanical facts are enunciated as if there is an absolute space to which they can be referred.
\\\hspace*{24mm}Henri Poincar\'e (1854-1912), Science and Hypothesis\\\hspace*{53mm}Chapter 6: Classical Mechanics 1902\end{quotation}

In Galilean and Poincar\'e relativity three-dimensional space does not exist as a physical reality, it is merely a mathematical \textit{convention}.

\begin{quotation}There is no entity 'physical space'; there is only the abstract space chosen by the physicist as a structure in which to plot phenomena; and some choices give simpler theorems than others (thus making the laws of nature look simpler).

The essence of scientific freedom is the right to come to conclusions which
differ from those of the majority.\\\hspace*{44mm}Edward Arthur Milne (1896-1950) [1951]\end{quotation}

Neither Einstein 1905, nor Minkowski in 1908, made such explicit and clear statements about the relativity of space, about a choice of a rigid body.

\begin{quotation}Henceforth space by itself, and time by itself, are doomed to fade away into mere shadows and only a kind of union of two will preserve an independent reality.\\\hspace*{70mm}Hermann Minkowski 1908\end{quotation}

The Minkowski `union of two' could suggest incorrectly the uniqueness of space, and spacetime as a Cartesian product of space and time. The primary concern of the theory of relativity is the necessity of the relativity of \textit{space}: that there are \textit{many} spaces. This notion of the relativity of space is of metric-independent, and requires no concept of simultaneity. This was also observed by Ruggiero [2003], and by Arminjon and Reifler [2010 \S 4 Discussion ii) on page 10].

The relativity of time is not a primary concern. We should not insist on the necessity of the relativity of time, because the only relative concept of time is the metric dependent proper time. Nature allows other metric-free conventions of simultaneity, such as radio-simultaneity, etc.
For related explications we refer to [Poincar\'e 1902; Trautman 1970; Matolcsi 1994 Part I \S 3; Selleri 2010].

In spite of the above assertions of Galileo and Poincar\'e concerning the necessity of \textit{many non-physical spaces}, it seems that the majority of the present-day scientific community believe in the existence of a unique three-dimensional physical space. In some publications the word `space' is always used in the singular, understood as unique and therefore as a physical concept that one can experience in Nature. For example Jammer's monograph [1954] entitled `Concepts of Space', avoids the plural `spaces'. The unique space was exactly the point of view of Aristotel in ancient Greece, however the Galilean revolution of \textit{many relative spaces,} each three-dimensional space as merely a mathematical convention, is still not widely accepted more than 400 years later!

\begin{center}\begin{tabular}{c|cc}&Space&Spacetime\\[3pt]\hline\\[-6pt]
\bt Ancient Greeks and\\some present Scientific\\community\et&\bt Space is unique\\and it has\\physical reality\et&\bt Spacetime\\is non-physical\\mathematical\\abstraction.\\Not an essential\\part of Nature\et\\\vspace{2mm}\\\bt Galileo 1632\\Poincar\'e 1902\et&\bt Space does not exist\\in Nature.\\There are many\\mathematical spaces\\as conventions\et&\bt Four-dimensional\\spacetime\\is physical\\reality\et\end{tabular}\end{center}

\section{Galilean relativity of spaces}\hspace{5mm} In 1632 Galileo Galilei observed that to be in the same place is \textit{relative}, \ie a subjective concept, not objective. Place is observer-dependent. Galileo implicitly (conceptually) introduced a four-dimensional \textit{physical} space-time of absolute events, after the experimentally confirmed observation that it is impossible to detect the motion of a boat without a choice of external reference system.

If the concept of place in a three-dimensional space needs an artificial choice of some physically irrelevant reference system, then three-dimensional space is an illusion. Different reference systems yield different three-dimensional spaces, and the only objective physical-arena is four-dimensional space-time (Galilean space-time or Minkowski space-time). Three-dimensional space is a \textit{mathematical convention} that depends on a subjective choice of reference system. According to Galileo: there are as many three-dimensional spaces as there are reference systems, \ie there does not exist any `unique physical space'. Therefore space-time must not be seen as it was by the Aristotelan Greeks: an Earth-space moving in time. Aristotle has only one unique observer: the Earth. Galilean space-time allows infinite number of observers.

Galilean space-time is a fiber-bundle (fiber is simultaneity submanifold) over one-dimensional time, without any preferred space [Trautman 1970]. Trautman claims that each fiber over a time-moment is `isomorphic to Euclidean 3-space $\Real^3$', that one can interpret a fiber over time as (isomorphic to) a physical space of places. This is not the case! Each fiber is a set of simultaneous \textbf{events}, and not a set of places in a `physical' space! There is no space concept within Galilean physical space-time, because the concept of the space needs an artificial \textit{choice} of the reference system. Galilean space-time is \textit{not} the cartesian product of time with some fixed space, because there does not exist a privileged space among the many spaces. There is not just a single space, there are infinite \textit{many} spaces.

If some reference system is chosen, Earth or Sun?, then the corresponding space of \textit{this} massive body is \textit{not} a fiber in space-time, but it is rather a quotient-space = space-time/material-body,
\begin{gather}\text{Space}\;\equiv\;\frac{\text{Space-time}}{\text{material body}}\qquad
\text{Time}\;\equiv\;\frac{\text{Space-time}}{\text{Convention of simultaneity}},\label{Space}\\
\text{Proper-Time}\;\equiv\;\frac{\text{Space-time}}{\text{Metric simultaneity of material body}}.\label{proper}\end{gather}

In our interpretation of Galileo Galilei: \textit{physical} reality is a four-dimensional space-time of \textbf{events}. Time can never "stop", and the choice of three-dimensional space is no more than a \textit{mathematical} convenience. The name \textit{space-time}, introduced by Hermann Minkowski in 1908, is misleading, suggesting incorrectly that this concept is derived from two primitive concepts of `space' and `time'. It is just the opposite, the most primitive concept is the Galilean space-time of events, and space is a \textit{derived} concept that needs an artificial \textit{choice} of massive body, e.g. Earth or Sun, as a reference system \eqref{Space}. But any such choice is irrelevant for physical phenomena, it is no more then for example a convenience for a computer program.

The Galilean four-dimen\-sio\-nal space-time does not possess an invertible metric tensor. The Minkowski version of Poincar\'e's and Einstein's special relativity added an invertible metric tensor, the Minkowski metric, to Galilean space-time.

\begin{figure}[h!]
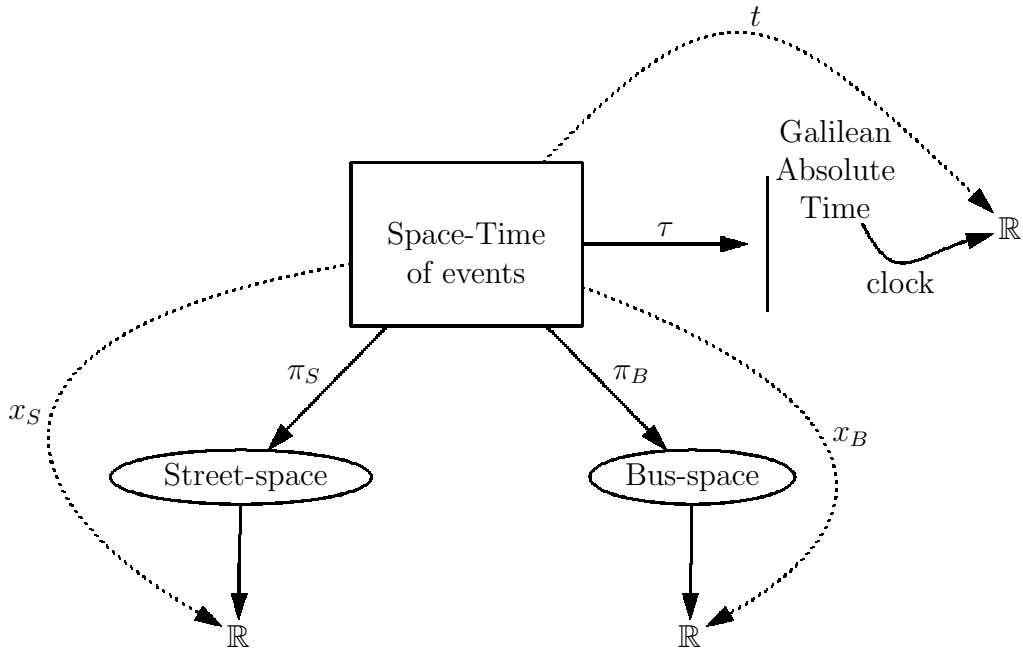
\Draw(1mm,1mm)
\RectNode(x)(--

\quad Space-Time\quad
\quad of events\quad
--)
\LineAt(40,-9,40,9)
\Move(-70,-40)\OvalNode(y)(-- Street-space--)\Move(60,0)\OvalNode(z)(--Bus-space--)
\MoveTo(49,10)\Node(w)(--Galilean
Absolute
Time--)
\MoveTo(38,0)\Node(u)(----)
\ArrowHeads(1)\Edge(x,u)\EdgeLabel(--$\tau$--)\Edge(x,y)\EdgeLabel[+](--$\pi_S$--)
\Edge(x,z)\EdgeLabel(--$\pi_B$--)
\MoveTo(70,0)\Node(a)(--$\Real$--)
\CurvedEdgeSpec(-45,0.4,35,0.3)\CurvedEdge(w,a)\EdgeLabel[+](--clock--)
\MoveTo(-30,-50)\Node(b)(--$\Real$--)\Edge(y,b)
\MoveTo(30,-50)\Node(c)(--$\Real$--)\Edge(z,c)
\EdgeSpec(D)\CurvedEdgeSpec(45,0.5,-45,0.5)\CurvedEdge(x,a)\EdgeLabel(--$t$--)
\EdgeSpec(D)\CurvedEdgeSpec(-50,0.9,90,0.5)\CurvedEdge(x,b)\EdgeLabel[+](--$x_S$--)
\EdgeSpec(D)\CurvedEdgeSpec(40,0.6,-90,0.5)\CurvedEdge(x,c)\EdgeLabel(--$x_B$--)
\EndDraw\caption{Two-body system, \{Street, Bus\}. To be in the same place is relative. Quotient $B$-space is different from quotient $S$-space.\label{M3}}\end{figure}

Galilean relativity postulates an absolute \textit{simultaneity relation}, denoted by $\tau$ on Figure \ref{M3}. Composed with a clock-function it gives a coordinate of spacetime of events,
\begin{gather}t=\text{clock}\circ\tau.\end{gather}
There is no need for another clock $t'=t.$ Absolute simultaneity is compatible with Einstein and Minkowski special relativity where it can be identified as just one among many different conventions of synchronization, such as for example the radio-synchronization which gives simultaneity that is metric-free [Marinov 1975; de Abreu and Guerra 2005].

\begin{odef}[Place] Each reference system is completely defined in terms of an equivalence \textit{relation} on events being in the \textit{same} place.

Thus every observer-monad field, $V\in\der\cF,$ gives rise to a surjective projection $\pi_V$ from four-dimensional space-time of events, onto a three-dimensional relative quotient $V$-space of places. Two space-time events, $e_1$ and $e_2,$ are in the same \textbf{place} for a $\pi$-observer if and only if, $\pi(e_1)=\pi(e_2).$\end{odef}

\begin{oexm} We must see how two reference systems, say a bus $B$ and a street $S,$ in a mutual motion, are distinguished within the space-time of \textbf{events}. Let us denote a street by $\pi_S$-system, and a bus by $\pi_B$-system. Lets illustrate the relativity of space in terms of the following list of three events:
\begin{center}\begin{tabular}[h]{ll}$e_1=$&bus start from the bus stop `Metro'\\$e_2=$&bus almost arrive to the next bus stop `Center'\\$e_3=$&late passenger arrived to the bus stop `Metro'\\\hline\end{tabular}\end{center}

From the point of view of the bus driver, this is the $\pi_B$-system, the driver is in the same $B$-place inside of the bus, bus is at $\pi_B$-`rest':
\begin{gather}\pi_B(e_1)=\pi_B(e_2),\quad\text{but}\quad \pi_B(e_3)\neq \pi_B(e_1).\label{placeB}\end{gather}\end{oexm}

From the point of view of the crowd standing on the street, the street is the $\pi_S$-system:
\begin{gather}\pi_S(e_1)=\pi_S(e_3),\quad\text{but}\quad \pi_S(e_2)\neq \pi_S(e_1).\label{placeS}\end{gather}

\begin{oexm}Another example is a space of Sun and a space of Earth (Copernicus versus Ptolemy). The events are
\begin{center}\begin{tabular}[h]{ll}$e_1=$& Greg born (in Long Beach in July)\\$e_2=$&Bill born (in Long Beach in January)\\$e_3=$&Jamie born (in Washington in July)\\\hline\end{tabular}\end{center}
Were any of them, Greg, Bill, Jamie, born `in the same place'?\end{oexm}

\section{Conclusion:\\Galileo Galilei still not understood}\hspace{5mm}
Presented here opinion that relative motion is coordinate-free, but must be understood as relative motion of material bodies with respect to each other or with respect to material reference system, is often attributed to Gottfried Wilhelm Leibniz (1646-1716) or to Ernst Mach (1838-1916). However we are sure that must be in the first place attributed to Galileo Galilei (1564-1642).

In 1911 Langevin considered that acceleration must have an absolute meaning, independent of the reference system, independent of the choice of space. However if the concept of the relativity of velocity is not accepted a priori as an explicit function of the artificial material reference system, \ie if the  relative velocity is not accepted as a binary or ternary function of material reference systems, then we must not yet talk about acceleration.

If relative velocity is reference-system-dependent, then how can we be sure a priori that a change of velocity, the covariant derivative of velocity, that must involve the covariant derivative of any reference system,  be reference-system-free, \ie be absolute?

We conclude that the Galilean relativity of space of 1632 is not yet understood nor accepted by the scientific community in XXI century.

Science should be based on dissent. But as science becomes publicly funded, ideas become entrenched, and science becomes dogmatic. Textbooks extort only one unique absolute truth. Consensus, not dissent, is considered to be a good way to progress. Alternative ideas are derided, and not heard, frequently not accepted for publication by the referee system.


\end{document}